\documentclass{appolb}
\usepackage{epsfig}
\begin{document}
\eqsec 
 % uncomment this line to get equations numbered by (sec.num)
\title{ELECTRONIC PROPERTIES OF RANDOM POLYMERS:\\
MODELLING OPTICAL SPECTRA OF MELANINS
}
\author{Kinga Bochenek 
\and
Ewa Gudowska-Nowak
\address{Marian Smoluchowski Institute of Physics,\\
 Jagellonian University, 30-059
Krak\'ow, Poland\\}
E-mail: kinga@indigo.if.uj.edu.pl, gudowska@th.if.uj.edu.pl}
\maketitle
\newcommand{\ben}{\begin{eqnarray}}
\newcommand{\een}{\end{eqnarray}}
\newcommand{\be}{\begin{eqnarray}}
\newcommand{\ee}{\end{eqnarray}}
\newcommand{\ba}{\begin{eqnarray*}}
\newcommand{\ea}{\end{eqnarray*}}
\newcommand{\der}[2]{\frac{\partial #1}{\partial #2}}
\newcommand\del{\partial}
\newcommand{\tr}{{\rm tr}}
\newcommand{\no}{\nonumber}
\newcommand{\mbold}[1]{{\bf #1}}

\begin{abstract}
Melanins are a group of complex pigments of biological origin, widely spread in all species
from fungi to man. Among diverse types of melanins, the human melanins,  {\em eumelanins},
are brown or black nitrogen-containing pigments, mostly known for their  photoprotective properties 
in human skin.
We have undertaken theoretical studies aimed to understand  absorption 
spectra of {\em eumelanins} and their chemical precursors. 
The structure of the biopigment is poorly defined, although it is believed
 to be composed of  cross-linked
heteropolymers based on indolequinones.
As a basic model of the {\em eumelanin} structure, we have chosen
pentamers containing hydroquinones
(HQ) and/or 5,6-indolequinones (IQ) and/or semi\-qui\-nones (SQ) often 
listed as structural melanin monomers.
 The {\em eumelanin} oligomers have been constructed as random compositions of
 basic monomers and opitimized for the energy of bonding.
 Absorption spectra of model assemblies have been calculated 
within the semiempirical intermediate neglect of differential overlap (INDO) approximation.
Model spectrum of eumelanin
has been further obtained by sum of independent spectra of singular polymers. 
By comparison with experimental data it is shown that the INDO/CI method
manages to reproduce well characteristic properties of experimental spectrum
of synthetic {\em eumelanins}. 

\end{abstract}
  
\section{Introduction}
Melanins are ubiquitous natural pigments formed by oxidation and polymerization of catechols
\cite{PROTA,SARNA}.  
Black or dark brown nitrogen-containing melanins are classified as {\it 
eumelanins} distinctly from  {\it pheomelanins} - whose lighter coloring originates from
intervention of cysteine in the process of melanogenesis, and {\it allomelanins} - which
originate from nitrogen-free precursors and are typical of plants and microorganisms.
The {\it melanogens} - uncoloured precursors of melanins of the animal kingdom - are
diphenols derived from phenylalanine and from tyrosine. 
The  melanin pigments are amorphous  substances of remarkable ability to absorb 
almost indiscriminately near-infrared, visible and ultraviolet radiation \cite{SARNA,PROTA1}. 
Exploiting these properties of melanins has proven to be a complex task, since despite 
significant experimental effort on a variety of natural and synthetic melanins, the
molecular structure of the pigment and its organization remains still unknown.
Poorly understood are also functional properties of natural melanin pigments which call for
attention because of their 
 both - photoprotective and photosensitizing properties. \\ \\
\noindent In this paper we will focus on hypothesis which concludes on spectroscopic properties of
melanins assuming their polymeric amorphous structure. Our studies concern {\em eumelanins}
which are known as
 complex broadband absorbers, frequently subject to excitation in its native environments.
 Eumelanin is mainly derived from 5,6-dihydroxyindole (DHI; for simplicity named also hydroquinone, HQ) and 5,6-dihydroxyindole-2-carboxylic
 acid (DHICA) \cite{ITO}. The detailed structural properties of this polymer are still
 under study because of unclear covalent and ionic configurations of HQ and DHICA
 monomers within the natural eumelanin unit. It is possible, however, to produce 
 synthetically pure DHI or
 DHICA melanin. In this respect, theoretical studies presented in this paper relate
 to synthetic eumelanins that are free or almost free from the DHICA contribution in the
 structure.\\ 
 Absorption bands of eumelanins have been recently \cite{ZAJAC,SIMON1,SIMON2} shown to depend on
 different levels of aggregation reported in scanning electron microscopy studies.
 The proposed aggregation structure is generally viewed as a molecular
 system containing 3-4 oligomers, often referred to in literature \cite{SIMON1,SIMON2}
 as  the fundamental aggregate. The hypothetical pigment structure is then believed to be 
 assembled from these $\pi$ stacked, crosslinked units. \\ \\
 In an earlier attempt to explain electronic (and optical) properties of melanins ({\it
 cf.} Figure 1), a band
 model \cite{McGinnes,PROCTOR}
 viewing melanins as amorphous semiconductors has been proposed.
 In fact, melanins can be considered as mixtures of more or less similar polymers,
 apparently made up of different structural units linked through heterogenous
 non-hydrozable bonds. The latter are the result of the polymerization of indole-type
 rings in both, the quinone and hydroquinone oxidation state, randomly crosslinked and
 piled. Theoretical contributions on a speculative model \cite{HIG,PULLMAN} pictured
 eumelanins as a linear chain semiconducting polymer. 
  Extrapolating the
 bonding character of the lowest unoccupied molecular orbital (LUMO) of one particular dimer
 of $5,6$-indolequinone to the lowest conduction band of the infinite polymer,
 Pullman and Pullman \cite{PULLMAN} 
 have pointed out the tendency of such a melanin model to be an electron acceptor
 that could explain trapping of free radicals characteristic for the natural and synthetic pigments.\\ 
 More recent studies along similar lines have been performed by Galv\~ao and Caldas
 \cite{Cal1,Cal2,Cal3}. Using the H\"uckel $\pi$-electron approximation
 and  the parametrization of Pullman \cite{PULLMAN},  the authors  studied the electronic
 structure of a family of ideal ordered polymers arising from the indolequinone in
 different redox states. It has been shown that the direction of polymerization 
 begins to
 emerge  as the length of the polymer increases. The authors pointed out also that the
 redox state of the melanin units played a role in its band structure, as {\it e.g.} the polymer
 built from 5,6-dihydroxyindole units exhibited larger gaps and narrower bands,
  whereas
 finite chains of semiquinone units exhibited bonding character of their LUMO.
 \begin{figure}[b]
\centerline{\epsfxsize=10.0truecm\epsfbox{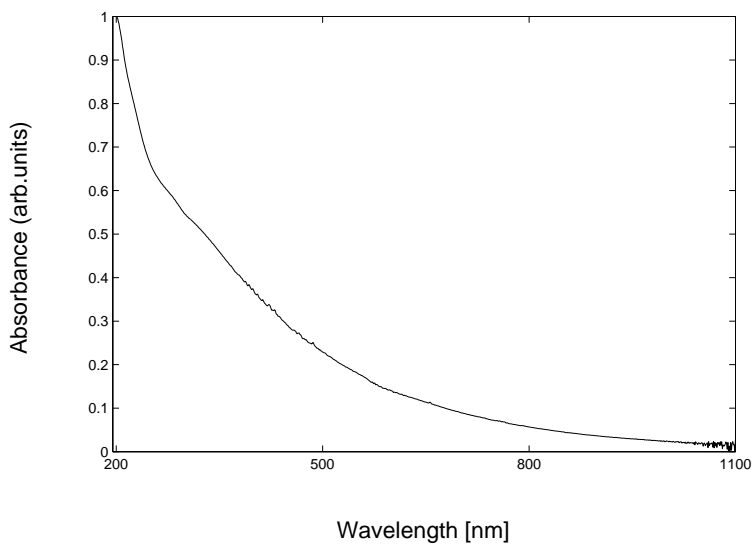}}
\caption[]{Experimental steady state absorption spectrum of synthetic DOPA 
melanin recorded using a UV-Vis spectrophotometer with an accuracy of 2nm. 
}
\label{fi1}
\end{figure}
 
Some limitation of this approach is the assumption of planarity of melanin polymers,
which in most cases does not need to be the case.
In fact, structural modelling of eumelanins based on comparing calculated data of the
reduced structure factor and radial distribution function for limited random network
models with the experimental X-ray scattering data \cite{CHENG} suggest that planar
models consisting of undistorted long chains are not appropriate structures for amorphous
melanins.\\ \\ 
\noindent We have thus considered  the role of conformational variations of eumelanins' oligomers
in modifying and modulation of their light-absorption properties. According to the existing
evidence \cite{PROTA,SARNA,PROTA1,YOUNG,ZHANG} that the 5,6-indolequinone (IQ) or the reduced forms,
 semiquinone
(SQ) and hydroquinone (HQ) compose the major part of the active  pigment, 
these three basic units ({\it cf.} Figure \ref{fi1}) in their neutral forms have been used to build up model
 polymeric structures.
 
 \begin{figure}[b]
\centerline{\epsfysize=8.0truecm\epsfbox{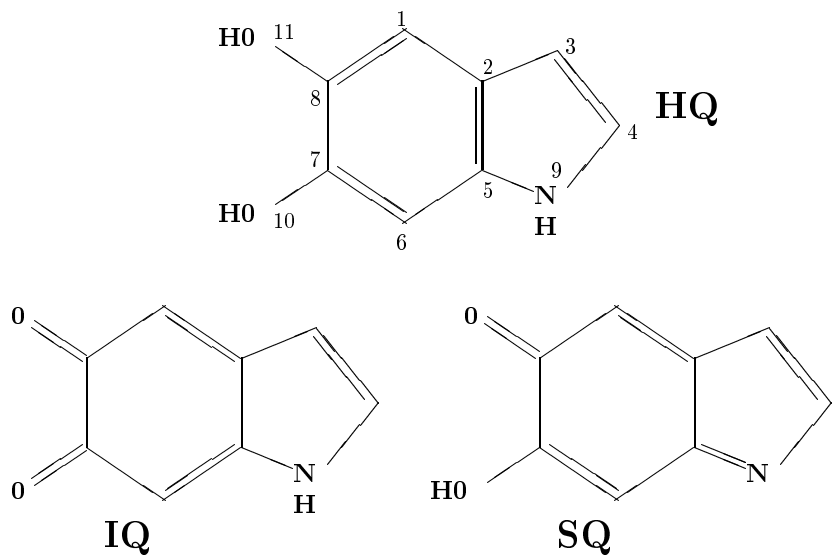}}
\caption[]{Schematic monomer structures used in this work
and labelling of active sites: {\bf HQ}- hydroquinone; {\bf IQ}-
5,6-indolequinone; {\bf SQ}- semiquinone. 
}
\label{fi2}
\end{figure}
 
When focusing on oligomeric models of eumelanins, we have assumed in consistency
 with the
wide-angle X-ray diffraction measurements of synthetic eumelanin samples and STM images of
monolayers  of synthetic eumelanins on graphite \cite{ZAJAC,CHENG},
that  a fundamental building block of the polymer is a structure consisting of several 
(4-10) 5,6-indolequinone residues composed in $\pi$-system molecular stacks. Therefore, as a starting model unit 
for
a  higher order polymeric structure of eumelanin a pentamer containing HQ, IQ or SQ
 molecules has been chosen 
and subsequently used to analyze randomly created structures of
 various conformations. Pentamer structures with various proportions of HQ, IQ and SQ
 content have been constructed from geometrically optimized monomers 
 and further analyzed for their spectroscopic properties. 
 All oligomer models have been  built by using 
"Random Polymer Builder" module of the Cerius$^{2}$/ MSI packet \cite{PROG}.\\

In assessing electronic spectral properties of modelled molecular structures, we report here results from the  spectroscopic ZINDO/s calculations developed by Zerner \cite{ZERNER1}
and co-workers. The method is based on re-parametrization of the semi-empirical INDO 
(Intermediate Neglect of Differential Overlap) approach that includes monoatomic differential overlap
for one-centre integrals ({\it i.e.} integrals involving basis functions on the same atom).
 These calculations yield absorption maxima which establish that the
differences in mixing ratios of various  monomeric units in an overall polymeric structure
of eumelanins can influence and alter absorption spectrum of the oligomer.

\section{Computational models}
\label{models}
In order to pinpoint the factors controlling the spectra of eumelanin polymers, the model
subunits have been treated at several levels of structural elaborations.
As basic components 
 of  eumelanins  5,6-dihydroxyindole (HQ), 
5,6-indolequinone (IQ) and semiquinone (SQ) (see Figure 2.) have been chosen. 
Since experimental data on the geometry of the molecules in Figure 2 are not available,
the bare skeletons of indole molecules have been edited by a molecular editor in the
InsightII/MSI module
 and
further geometrically optimized by use of either {\it ab initio} Gaussian-94
\cite{GAUSS} or
semiempirical consistent
valence forcefield (CVFF) method. In both cases the energies for conformation were minimized until
the convergence  was achieved (the maximum derivative was less than $0.0001$ 
kcal mol${}^{-1} \AA^{-1}$).
The augmented CVFF program has been developed for material science applications
\cite{DAUBER}
 and is
considered a standard forcefield to be used for organic molecules and polymers.
 As a default forcefield in Discover 95.0 molecular dynamics simulation package, 
 it has been extensively used over the past 
 years and is usually considered \cite{DAUBER,HAGLER,PROG} well tested and characterized.
 It is primarily intended for studies of structures and binding energies,
  although is
 known also to predict satisfactorily
 frequencies in vibrational spectra as well as conformational energies.\\ 
Structures optimized by CVFF method have been further used as input parameters for
 spectroscopic
ZINDO \cite{ZERNER1} calculations. The method circumvents cumbersome and 
time-consuming {\it ab initio} treatment of complex systems
and 
 is one of commonly used semiempirical molecular orbital (MO) methods. Over the last decade it
  has been succesfully applied in the study of bulk solids and defects in semiconductors \cite{SOLID}.
  The INDO electronic calculations are based on an all-valence-electron, self consistent
   field (SCF)
  molecular orbital procedure with the configuration interaction (CI).
 Within the general MO method, the Fock matrix 
can be represented as
\be
F_{\mu\nu}=h_{\mu\nu}+\sum^N_{\lambda,\sigma}P_{\lambda,\sigma}[(\mu\nu|\lambda\sigma)-\frac{1}{2}(\mu\lambda|\nu\sigma)]
\label{fock}
\ee
where $h_{\mu\nu}$ stand for the core integral
\be
h_{\mu\nu}=\int \phi_{\mu}(1) H_{core} \phi_{\nu}(1) d\tau_1,
\ee
 $(\mu\nu|\lambda\sigma),(\mu\lambda|\nu\sigma)$
represent two-electron Coulomb integral and the exchange integral, respectively, 
\be
(\mu\nu|\lambda\sigma) =\int\int\phi_{\mu}(1)\phi_{\nu}(1) r^{-1}_{12}
\phi_{\lambda}(2)\phi_{\sigma}(2) d\tau_1 d\tau_2
\label{two}
\ee
The electron density matrix $P_{\lambda\sigma}$ 
\be
P_{\lambda\sigma}=2\sum^{occ}_{i=1}c^*_{\lambda i}c_{\sigma i}
\ee  
contains molecular orbital
expansion coefficients $c_{\lambda i}$ of the $\Psi_i$ orbital formed as a
linear combination of basis functions $\phi_{\lambda}$.
The variation principle $
\frac{\partial E}{\partial c_{\mu i}} = 0$
applied to a closed-shell system
leads then to the {\sl Roothaan--Hall} equations:
\be
\sum^N_{\nu=1} (F_{\mu\nu}-\epsilon_i S_{\mu\nu})c_{i\nu} = 0
\label{root}
\ee
for the set of orbital energies $\epsilon_i$ and MO coefficients $c_{i\nu}$.
The elements of the matrix $S_{\mu\nu}$ are the AO overlap 
integrals $\int \phi^*_{\mu}(1)\phi_{\nu}(1) d\tau$.

Equation (\ref{root}) is solved iteratively until a self-consistency is reached
within the demanded accuracy. 
The final SCF solution  yields desired MOs $\Phi_i$ and their
orbital energies $\epsilon_i$. The ground state electron configuration is produced by 
filling the orbitals with all electrons in the order of increasing energy.

The SCF calculations done with a set of $M$ basis functions require the computation of $M^4$ matrix elements. In order to make a treatment of large molecules possible, one has to reduce the complexity by either replacing the effect of inner core electrons by effective (pseudo-) potentials or by applying
the semi-empirical SCF method which neglects most of the matrix elements
$(\mu\nu|\lambda\sigma)$ and parametrize the remaining ones.
 The term "intermediate neglect" points to the retention of one-center-two-electron exchange integrals
  $(\mu\nu|\mu\nu)$.
Within the Zerner's version of INDO (ZINDO) approximation \cite{ZERNER1}, basis orbitals $\phi_i$ 
are envisioned to be strongly orthogonal and they obey the relations
\be
(\mu^A\nu^B|\lambda^C\sigma^D) & \equiv & \delta_{AB}\delta_{CD}\int\int\phi_{\mu}^A(1)\phi_{\nu}^B(1) r^{-1}_{12}
\phi_{\lambda}^C(2)\phi_{\sigma}^D(2) d\tau_1 d\tau_2\nonumber\\ 
& = &(\mu^A\nu^A|\lambda^A\sigma^A)\qquad A = C\nonumber\\
& = &(\mu^A\nu^A|\lambda^C\sigma^C)\qquad A \neq C
\label{zasada}
\ee
where $\phi_{\mu}^A$ is the atomic orbital centered on atom $A$.
In order to maintain rotational invariance, two--center integrals ($A\neq C$) are 
evaluated over atomic orbitals  $\tilde{\phi}_{\mu}^A$ that are $s$ symmetric but have the 
same exponents and expansion coefficients as  $\phi_{\mu}^A$.
The one-center
core integrals $h_{\mu\mu}$ are obtained from ionization potentials and
the resonance integrals (the two-center $h_{\mu\nu}$ integrals) 
 are purely empirical parameters set to reproduce experimental spectra.
The method as employed herein leads to the
ground electronic states  obtained as closed-shell molecular orbital
wave functions in the restricted
Hartree-Fock (RHF) framework. In the next step, low lying excited states were approximated by configuration interactions
(CI) among configurations generated as single excitations from RHF ground state. The CI 
method included the highest 15 occupied and lowest 15 unoccupied molecular
 orbitals. \footnote{Expanding the active basis of the CI space have not resulted
 in significant differences in modelled absorption spectra, therefore all
 oligomeric structures have been treated within the same CI-conformity.}  \\ \\ 
 All polymers were arbitrarily chosen as pentamers built by 
use of the "Random Polymer Builder" module of the MSI packet\cite{PROG}. Four different groups
 of polymers: homogenous polyHQ, polySQ, polyIQ ({\it cf.} Figure 3) and heteropolymers 
 (~{\it i.e.} pentamers built up from basic units with random ratios of  molecules HQ:IQ:SQ) 
were investigated. The polymer structures were formed by defining all possible pairs of "head" 
and "tail" positions
out of 5 labeled centers (see Figure 2) located on a monomer. Structural configuration models of
pentamers have been then developed along the polymerization direction in which "heads" were joined
with "tails" of subsequent monomers. Resulting pentamer structures ({\it cf.} Figure 3)
can be either linear and flat, nonplanar linear or forked structures, respectively.
Despite their various conformational structures, all three groups are characterized by a
mean dimension (measured as a maximum observable distance between non-hydrogen atoms in a
model pentamer form) of about $2nm$ and molecular weight about 720-740, {\it i.e.} with
$MW<1000$. If assuming that such units are basic chromophores in the eumelanin structure, 
the models predict that there are about of million of such pentamers in a melanin subunit
of an average diameter $150-200nm$ \cite{KTOS,CHENG}.\\
 Heteropolymeric structures have been 
created either with equal
probabilities of choosing IQ, SQ or HQ molecule at each step of polymerization,
or by using IQ and HQ monomers only, {\it i.e} with the HQ:IQ:SQ probability proportion
taken as 3:1:0. The latter choice switches off the reduced form of IQ, the
component SQ, that is mainly responsible for
formation of free radicals produced during melanin synthesis \cite{SARNA}.
 The planar and non-planar conformations were
analyzed seeking for their effect on absorption spectra. Models of  planar 
linear polymers (each monomer had no more than 2 bonds), planar forked polymers 
and non-planar linear 
polymers (see Figure 3.) have been investigated for their  absorption spectra, calculated
solely for neutral structures by using ZINDO-CI method.\\
 
 \begin{figure}[b]
\centerline{\epsfxsize=10.0truecm\epsfbox{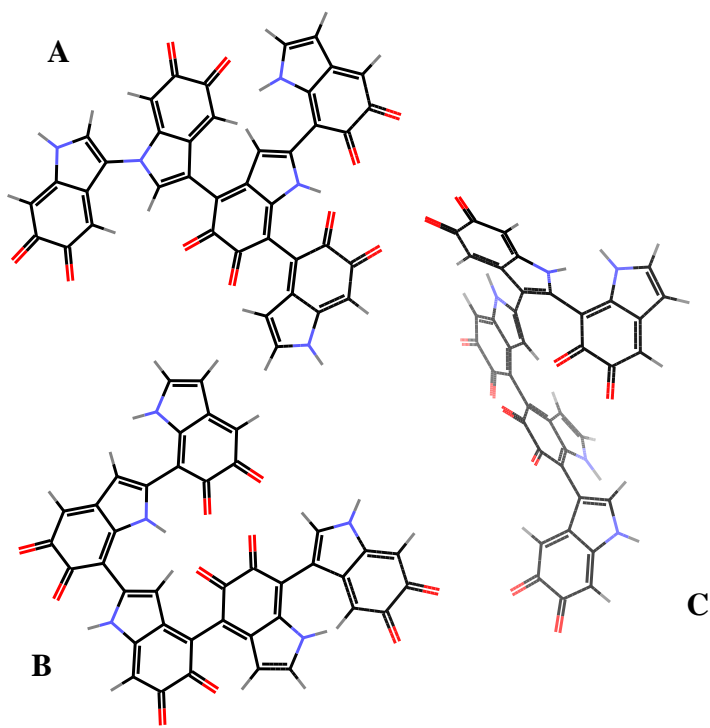}}
\caption[]{Examples of polymers used in calculations: A - planar and 
                  forked, B - planar and linear, C - nonplanar,linear. 
}
\label{fi3}
\end{figure}

Dynamic interactions between the model basic "molecule" and the surrounding medium  (composed
of the set of randomly distribured similar oligomeric units) can be a source of
fluctuations in the electronic gap energy between the ground and excited state.
To account for these fluctuations and their effect on the system response to a
radiation field, we have presented calculated spectra adopting a Brownian oscillator model \cite{mukamel} that
relates variability in the electronic gap to displacements in some dynamical
variable $x$ representing the system under consideration.
In the linear regime (weak coupling to the perturbing external radiation field),
the fluctuation-dissipation theorem relates the system's response function to
the derivative of the correlation function $C(t)=<x(t)x(t')>$ with brackets
representing an equilibrium average. In the Brownian oscillator model the
equilibrium states are defined by incorporating a model of the bath represented
by a Gaussian  stochastic force $f(t)$ acting on  a variable $x$:
\ben
m_i\ddot{x}_i(t)+m_i
\omega_i^2x_i(t)+m_i\int^t_{-\infty}d\tau\gamma_i(t-\tau)\dot{x}_i(\tau)=f_i(t)+F_i(t)
\een
where $m_i$, $F_i(t)$  stand for the "mass" of a mode $x_i$ and  
an external driving force, respectively. The correlation of the random force is
related to the dissipative term  (friction), {\it i.e.}
$<f_i(t)f_k(t')>=\delta_{ik}2m_ik_BT\gamma_i(t-t')$. The variation of the
friction term with time (or frequency) reflects the timescale of the thermal
motions of the bath. By assuming that the latter are very fast compared to the
oscillator motion, we accept the ohmic dissipation model with $\gamma_i$ being a
constant. In a strongly overdamped case ($\gamma_i>>2\omega_i$) and in a fast
modulation limit \cite{mukamel}, the absorption lineshape calculated as
\ben
\sigma_{abs}(\omega)=\frac{1}{\pi} Re\int^{\infty}_{0}dt\exp[i(\omega-\omega_{eg})t]\exp[-g(t)]
\een
with
\ben
g(t)\equiv\int^t_0 dt_2\int^{t_2}_0dt_1C(t_1)
\een
assumes a Lorentzian form
\ben
\sigma_{abs}(\omega)=\frac{\Gamma}{\pi[(\omega-\omega_{eg})^2+\Gamma^2]}
\een
with the line width $\Gamma$  proportional to the product of characteristic nuclear time
 scale (characteristic time scale of medium relaxation) and a coupling strength
 of the nuclear degrees of freedom to the electronic transition. 
 Without discussing the origin of the particular choice of the lineshape in more
details, we have thus arbitrarily adopted that the energy transitions calculated
for the model system were 
Lorentzian  enveloped and summed, weighted with an oscillator strength to simulate the 
overall absorption spectra. 
Differently chosen
  values of $\Gamma$ reflect then the degree of dissipation of absorbed energy
  {\it via} coupling  to the bath.

\section{ RESULTS}
\subsection{\bf Monomers}
Geometrical features of the ground state monomers used in this study have been compared with
optimized coordinates from {\it ab initio} Gaussian 94 calculations.
 In accordance with the
previous analysis by Bolivar-Marinez {\it et al.} \cite{Cal3},  various methods of optimization
result in small variation of main bond lengths and bond and dihedral angles
yielding  structures that are basically planar.  This feature is
further reflected in the major $\pi-\pi^*$ character of the HOMO-LUMO transition 
for all monomer
molecules.
To assess the reliability of our calculations based on CVFF method combined with ZINDO/s 
calculations in predicting spectral features of melanin's derivatives, we have calculated transition
energies for the bare skeleton of HQ (with each omitted substituent replaced by a hydrogen atom),
after optimization by either CVFF or PM3 method which has been used in a previous study \cite{Cal3}.
The first transition line calculated in ZINDO  was $306.5 nm$ ($296.6nm$) for the structure
optimized with CVFF (PM3) program, respectively. This has been compared with the first transition
line ($300.3 nm$) calculated for the "empirical" bare skeleton structure of tryptophan obtained
 from the Protein
Data Bank. 
\renewcommand{\arraystretch}{1.5} 
\begin{center}
\begin{minipage}{20cm}
\begin{tabular}{r|c|c}
monomer&Gaussian/ZINDO&CVFF/ZINDO\\
\hline
  & energy (osc. strength)& energy (osc. strength)\\
  & [nm]&[nm]\\
\hline\hline
HQ&307.2(0.05)&316.2(0.06)\\
  &228.0(0.82)&232.6(0.87)\\
\hline
IQ&464.5(0.12)&529.9(0.13)\\
&218.9(1.03)&225.3(0.52)\\
\hline
SQ&481.0(0.02)&856.4(0.03)\\
&217.4(0.94)&238.2(1.46)
\end{tabular}
\end{minipage}
\vskip 0.5truecm
{\small Table 1: Calculated excitation energies [nm] for optimized structures of melanin's
precursors. First row: first transition, second row: strongest transition. Oscillator strength are
given in paranthesis.} 
\end{center}
\vskip 0.5truecm
In all these cases, the first transition lines were predicted with oscillator strengths
of about
0.02. By inspection, all structures displayed also similar character of the calculated CI
coefficients and were comparable with the previous theoretical \cite{Cal3} and
experimental \cite{ZHANG} works related to melanin's monomers.\\
Results of Gaussian/ZINDO and CVFF/ZINDO calculations on neutral HQ, IQ and SQ monomers are presented 
in Table 2. The first optically active electronic transitions (first row entry
in Table 2)
and the strongest active electronic transition (second row entry in Table 2)
exhibit shifts towards longer wavelengths after CVFF structure optimization
which allows structures with some degree of non-planarity.\\ 
Reported transitions
are dominated by a few configurations involving frontier orbitals.
Inspection of the calculated CI coefficients reveals that about 70\% of the
first excited state for HQ monomer is accounted for by HOMO$\rightarrow$LU\-MO
excitation and about 15\% of  HOMO-1$\rightarrow$LUMO+1 excitation.
 The
strongest transition is composed of mixing among 2 highest occupied and 3 lowest
unoccupied orbitals. The pattern of the first excited state is essentially 
preserved for SQ and IQ
monomers where $>$96\% of excitation  comes from HOMO$\rightarrow$LUMO transition
with remaining excitations contributing with less than 1\%. \\ 
\subsection{\bf Oligomers and Polymers}
Figure 4 presents simulated optical absorption spectra of model polymers. The curves have been
obtained from the optical transition spectrum enveloped by normalized Lorentzians weighted by the
calculated oscillator strength; {\it i.e} energies ($\lambda_{i}$) and oscillator strengths ($s _{i}$)
obtained from calculations were converted to:
\ben
\sigma_{abs}(\lambda )=\sum_{i} \frac{s_{i}}{(\lambda - \lambda_{i})^{2} + 10^{2}}
\een
with the width of a line set up arbitrarily to 10 (in units of  $\lambda$).
The graphs refer to  spectra of different mixtures of polymers:
the first column displays spectra of the uniform
 mixture of polyHQ 
polymers ({\it i.e.}  pentamers composed of HQ monomers, only), second and third columns 
show absorbance spectra of similar uniform  mixtures of
  polyIQ and  polySQ, respectively. The fourth column presents spectra of
 heteropolymers composed with equal probabilities of choosing 
  either one of three various monomers ({\it i.e.} with HQ:IQ:SQ=1:1:1)
   at each step of the "polymerization process" leading to a basic pentamer structure.
 The last  column displays spectra of nonuniform polymers built  by use
 of the proportion HQ:IQ:SQ = 3:1:0 at the basic level of modelling. Rows in Figure 4 
 relate to spectra
  obtained for different conformations of polymers.
 The first row represents mixtures of  linear and planar polymers ({\it cf.} Figure 3), the second and third
 row  display
 spectra of linear nonplanar and planar forked polymers, respectively.
 
 \begin{figure}[b]
\centerline{\epsfxsize=10.0truecm\epsfbox{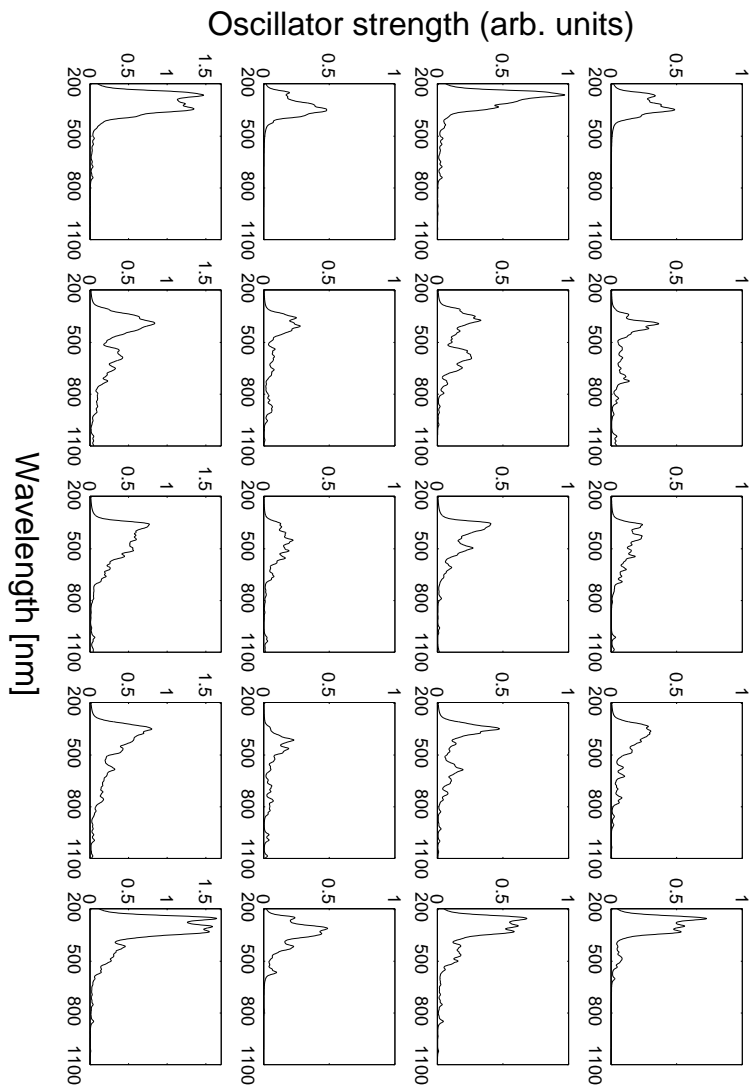}}
\caption[]{Theoretical absorption spectra for different mixtures of polymers:
first column - polyHQ, second column - polyIQ, third column - polySQ, fourth and fifth columns - heteropolymers 
(HQ:IQ:SQ = 1:1:1 and HQ:IQ:SQ = 3:1:0, respectively).
Rows refer to different conformational forms: first row - spectra of 15 
linear planar polymers, second row - 15 linear non planar polymers, third row - 15 forked 
planar polymers. Last row represents sums of spectra in each column (45
 polymers).
}
\label{fi4}
\end{figure}
 All spectra up to the last row in Figure 4 have been calculated as mixtures of 15 pentamers,
 whereas   the bottom row  shows sums of
simulated spectra for 45 polymers whose absorbance spectra are displayed in upper three rows
of each column. As an example, the third column and second row of Figure 4 displays
spectrum of 15 linear, nonplanar pentamers composed of SQ monomers, only. Accordingly, the inlet
in 5th column and 3rd row presents absorption spectrum of 15 "forking" pentamers composed of HQ and
IQ molecules taken in proportion 3:1 (with no contribution of SQ molecules). Sum of spectra for
various conformational structures of that type is displayed in the bottom right corner of Figure
4.\\
As can be inferred from Figure 4, spectra of polyHQ have dominant contributions within the
range $200-400nm$, whereas admixture of IQ or SQ components results in a  red shift
of the absorption lines. Conformation variations seem to result in varying intensity of
absorption, it is ambiguous however, to conclude on the overall effect of conformations on
spectra of model polymers.\\
 \begin{figure}[b]
\centerline{\epsfxsize=10.0truecm\epsfbox{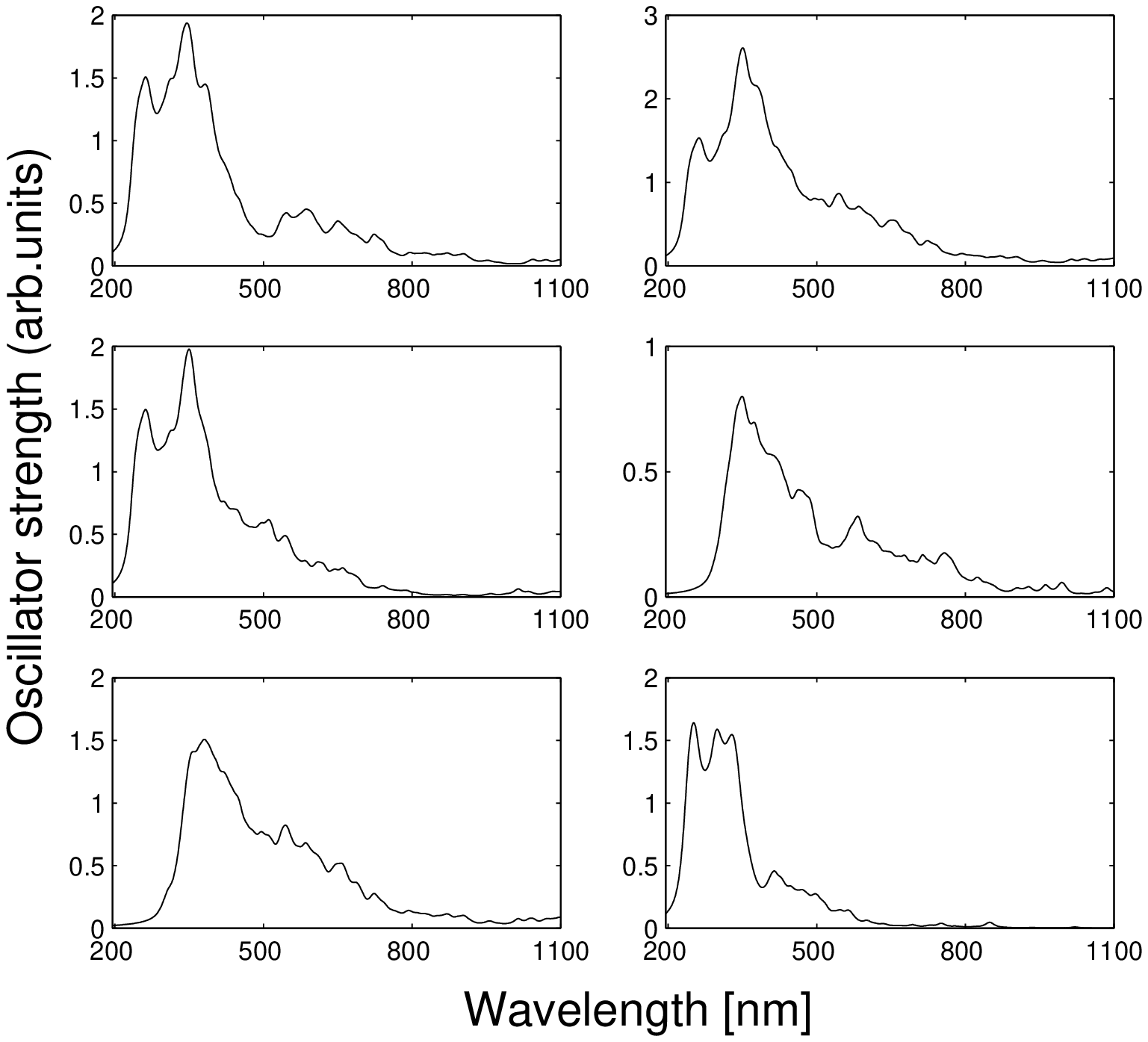}}
\caption[]{Simulated spectra for "higher order" mixtures of polymers: 
left column, top - 45 polyHQ + 45 polyIQ, below - 45 polyHQ + 45 polySQ, bottom
- 45 polyIQ + 45 polySQ.
Right column -  45 polyHQ + 45 polyIQ + 45 polySQ, below - 45 heteropolymers (HQ:IQ:SQ = 1:1:1),
bottom - 45 heteropolymers (HQ:IQ:SQ = 3:1:0).
See text for details.
}
\label{fi5}
\end{figure}

Figure 5 comprises model absorption spectra of "higher order" polymers:
in the top left corner spectra  of 45 polyHQ and 45 polyIQ oligomers are displayed,
second position in the first column represents spectra of 45 polyHQ and 45 polySQ, whereas
the bottom of first column contains simulated spectra for mixture of 45 polyIQ and 45
polySQ polymers. For comparison, spectra of mixture of three uniform oligomers (45 polyHQ +
45 polyIQ + 45 polySQ) have been displayed (second column and first row in Figure 5) along
with  45 heteropolymers HQ:IQ:SQ=1:1:1 (second row) and 45 nonuniform oligomers  
HQ:IQ:SQ = 3:1:0 (second column, third row). 
Note that all ``higher order'' spectra of polymers are constructed as direct
sums (mixtures) of spectra of pentamers and as such they do not take into
account interactions among those units.\\
 Simulated  optical absorption  spectra suggest
that the main contribution to the absorption  in
the range of $500-800nm$ may be attributed to the presence of SQ molecules in oligomeric
structures. Shape of simulated spectrum of mixtures of nonuniform polymers  resembles experimental
spectrum
(Figure 1) for synthetic DOPA melanin. The maximum  intensity of modelled spectrum is,
however, shifted to about $300nm$, whereas the peak intensity in the experimental spectrum
is observed already at $200nm$.
It can be expected that additional modulation of spectrum may arise from some
environmental effects. For example,  presence of localized charges may be a source of
electrochromic shift in the spectrum with red or blue shifts in transition energies 
determined by the sign and location of charges \cite{HANSON,EWA}. Possible shifts in transition
energies of model polymers may be also induced by higher content of charged forms of IQ 
monomers in model structures.
In particular, 
 it has been demonstrated in the previous theoretical study by Bolivar-Marinez {\it et al.} \cite{Cal3}
 that negative ions of IQ, SQ and HQ molecules (-1) begin to absorb at about 1.0eV.
 This effect is responsible, however, for the red shift in simulated 
  optical absorption spectra of negatively charged monomers. 
  \begin{figure}[b]
\centerline{\epsfxsize=10.0truecm\epsfbox{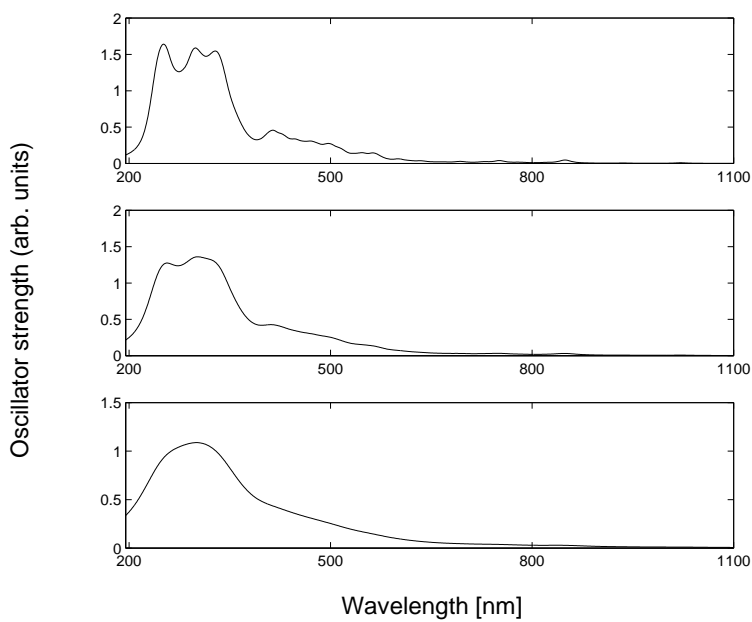}}
\caption[]{Effect of assumed line width on spectra of 45 heteropolymers
(built from HQ and IQ monomers only). Starting from the top, 
the values of Lorentzian's widths have been set up to 10, 20 and 40 (in units of the wavelength),
respectively. 
}
\label{fi6}
\end{figure}

 Calculated optical spectra smoothen also and shift when assuming
stronger dissipation of excitation energies in the system, resulting  in a broader width of lines
(see Figure 6). In the case of heterogeneous oligomers models (HQ:IQ:SQ = 3:1:0) fourfold higher line width
 parameter than originally chosen leads to a single humped spectrum with a broad, structureless tail.
 The effect, however,  does not alter significantly absorbance in the $200nm$
 region. \\
 Synthetic {\em eumelanin}  whose properties have been modelled in this study lacks 
 a protein coat and therefore its  optical spectrum does not match the
 natural pigment, especially in the region about $250-270nm$, where existing
 evidence \cite{KTOS} of spectral features has been attributed to absorption of 
 light by
 protein environment.

\section{CONCLUSIONS}
\vspace*{0.1cm}
The absorption spectra for model of {\em eumelanins} were calculated assuming melanin to be composed of
mixture  of various unit-polymers. All molecules were parametrized as
random pentamers of HQ, SQ and/or IQ. 
The idea of "mixed nature" of the pigment, composed of heterogeneous oligomers
 in a random, not necessarily well organized assemblies is compatible with experimental data. 
Simulated optical spectra manifest the effect of aggregation in oligomer conformation
resulting in broadening of the absorbance arm in the region of $500-800nm$.
Peak intensity of spectra can be correlated with the presence of neutral HQ monomers in model structures.
At higher level of aggregation (135 pentamers and more), further addition of a polymeric
 unit (a pentamer) to
the existing 
structure does not alter the character of the spectrum. Also, conformational deformations
of constituents (oligomers) have only a minor effect on the structure of modelled optical
spectra. Our future studies aim at understanding stabilisation of radicals in a
system of stacked oligomers and interpretation of factors responsible for
radical-trapping processes in eumelanins. 

\section{ACKNOWLEDGEMENTS}

The authors acknowledge many fruitful discussions with Prof. Tadeusz Sarna and Mr. Albert
Wielgus who has kindly provided the experimental spectrum for the synthetic DOPA melanin.
K.B. acknowledges hospitality of Prof. J. Brickmann and the group of Physical Chemistry II at the TU Darmstadt
where part of computational analysis has been performed. \\
This work has been supported by the Marian Smoluchowski Institute of Physics Research Grant
(1999-2002).  
\newpage
%{\bf REFERENCES.}
%\vspace*{-5mm}

\end{document}